\documentclass{moriond}

\def\be{\begin{equation}}
\def\ee{\end{equation}}
\def\bea{\begin{eqnarray}}
\def\eea{\end{eqnarray}}

\newcommand{\Photo}{}

\usepackage{graphicx}
\begin{document}
\vspace*{4cm}
\title{Higgs Boson with Multiple Jets}

\author{ Tuomas Hapola }

\address{Institute for Particle Physics Phenomenology, Department of Physics, \\
Durham University, DH1 3LE, United Kingdom}

\maketitle

\abstracts{
The High Energy Jets (HEJ) framework accounts for large logarithms arising from the wide angle hard gluon emissions. The resummation in HEJ is based on an approximation to allow fast evaluation for arbitrary multiplicity. The approximation is motivated by the high energy limit and as such captures the leading logarithmic corrections in large invariant mass between the partons. These corrections are important when there is a large separation in rapidity between the produced jets. This situation arises in a Higgs boson plus jets analysis. Furthermore, HEJ includes matching to full tree level accuracy up to four jets. This talk will introduce the HEJ framework and discuss advances in the formal accuracy attained within HEJ, and the application to predictions for the production of Higgs boson plus jets.
}

\section{Introduction}

The most characteristic aspect of analysis at hadron colliders is the impact of the abundant production of jets. Understanding high multiplicity jet production is vital for many Beyond Standard Model searches and precision QCD measurements. Moreover, jets can be a useful tool in the measurements trying to establish the properties of the Higgs boson.

In order to access the Higgs boson couplings, different production processes need to be separated. Weak boson fusion (WBF), where the Higgs boson is radiated from a t-channel W or Z boson, is characterised by the two tagging jets. This same topology can also arise from the higher order QCD corrections to gluon fusion (GF). In order to measure the WBF cross section, and get handle on the Higgs couplings to weak gauge bosons, the background from GF must be known. In addition, GF  can be used to study the $CP$-structure of the Higgs boson couplings to heavy fermions. The interference between GF and WBF is shown to be negligible \cite{Andersen:2007mp} making it justifiable to consider them separately.  Here we are interested in describing the Higgs plus two or more jets arising from GF. 

The main difference between the two processes is due to colour structures. In the GF case, the two jets are colour connected and effectively form a colour dipole. This is not the case for WBF, making these two processes to have a very different radiation pattern. This also suggests that GF is similar to W boson production in association with at least two jets. The dominant contributions to both processes are due to t-channel colour octet exchange.

Describing multi-jet processes is challenging for many reasons. For example, there are many different scales involved and the higher order perturbative corrections can be large. The High Energy Jets (HEJ) formalism provides an all order description of multi-jet processes which is motivated by the high energy limit. 

\section{High Energy Jets framework}

In the intermediate momentum region where the momentum transfer, $Q$, is much smaller than the centre of mass energy, but in the perturbative region, the production of a large number of partons is relevant. In this momentum range the series expansion parameter will be accompanied  by a large logarithm, and the effective expansion parameter can be of order one, $\alpha_s\log(s/Q) \sim \mathcal{O}(1)$. Thus one should take into account all orders in the effective expansion parameter.  A multitude of real emission channels will then cause the abundant production of partons. 

In order to do the resummation, a suitable limit is needed which allows $2 \to n$ matrix elements to be written in a simple form. In the high energy limit, or multi-regge kinematic (MRK) limit,
\begin{equation}
y_{1}\gg y_2\gg\cdots\gg y_n, \qquad |k_{\perp i}|\sim |k_{\perp j}|, \qquad \forall i,j,
\end{equation}
where $y_i, k_i$ are rapidities and momenta of the outgoing quarks and gluons, the QCD matrix elements can be written in a simple t-channel factorised form. This form is maintained also after the inclusion of virtual corrections to next-to-leading logarithmic accuracy in $s/t$. The t-channel factorisation is a consequence of gluon reggeazation and Regge theory which says that the asymptotic form of the amplitude is given by the t-channel exchange of the particle with the highest spin.
 
 The MRK limit restricts only the asymptotic form of the amplitude. The sub-leading logarithms are not systematically included in HEJ but the form of these sub-leading contributions is constrained by requiring Lorentz invariance and gauge invariance throughout all the phase space. The HEJ matrix elements for the  process $XY \to Xg\cdots gY$, $X,Y \in \{\rm{quark, gluon}\}$, can be written as\cite{Andersen:2009nu,Andersen:2011hs}
\begin{eqnarray}
\label{eq:m}
\overline{|\mathcal{M}_{XY \to Xg\cdots gY}|}^2 = \frac{1}{4(N_C^2-1)}||S_{XY\to XY}||^2\left( g^2 C_X\frac{1}{t_1} \right)\left( g^2 C_Y\frac{1}{t_{n-1}} \right) \nonumber \\
\prod_{i=1}^{n-2} \left( \frac{-g^2C_A}{t_i t_{i+1}}V^\mu(q_i,q_{i+1})V_\mu(q_i,q_{i+1}) \right)\prod_{j=i}^{n-2}\exp \left[ \mathbf{\omega}^0(q_j)(y_{j-1}-y_j)\right],
\end{eqnarray}
where $q_1=p_a-p_1$, $q_{i+1}=q_i-p_{i}$, $t_i=q_i^2$ and
\begin{equation}
||S_{XY\to XY}||^2=\sum_{\rm{helicities}}\left\vert j^\mu(p_1,p_a)j_\mu(p_n,p_b) \right\vert^2.
\end{equation}
The exponential terms encode the finite reminder left from the cancellation of real and virtual poles. The exact forms of the $\omega^0$ and the effective gluon emission vertex $V_{\mu}(p,q)$ are given in Ref. \cite{Andersen:2009nu}. The colour factor $C_{X(Y)}$ takes a value $C_A$ or $C_F$ depending if the $X(Y)$ is a gluon or a quark respectively. 

\begin{figure}[h]
\centering
\includegraphics[width=0.47\textwidth]{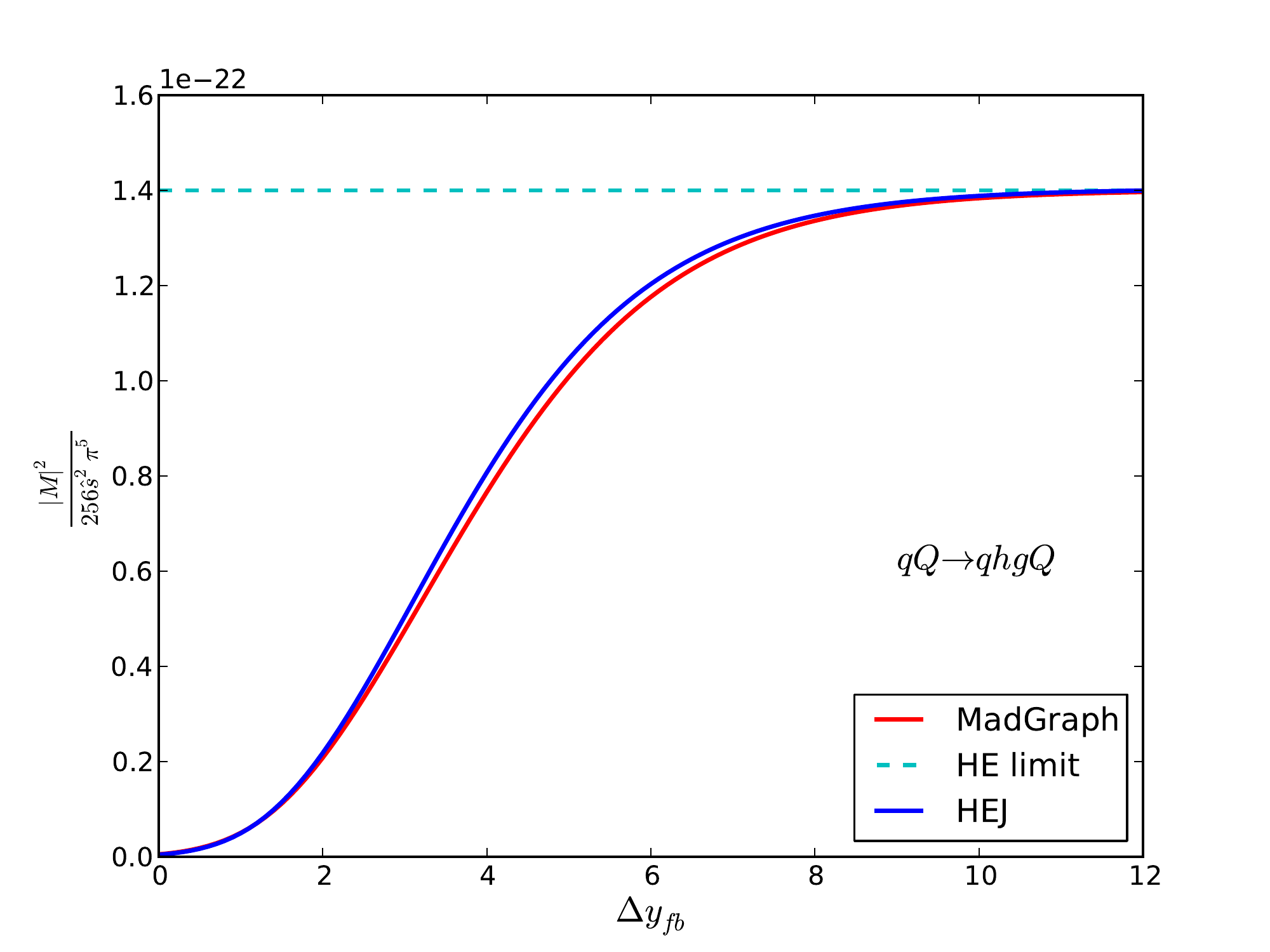}  \includegraphics[trim=0 -70 0 0,clip,width=0.47\textwidth]{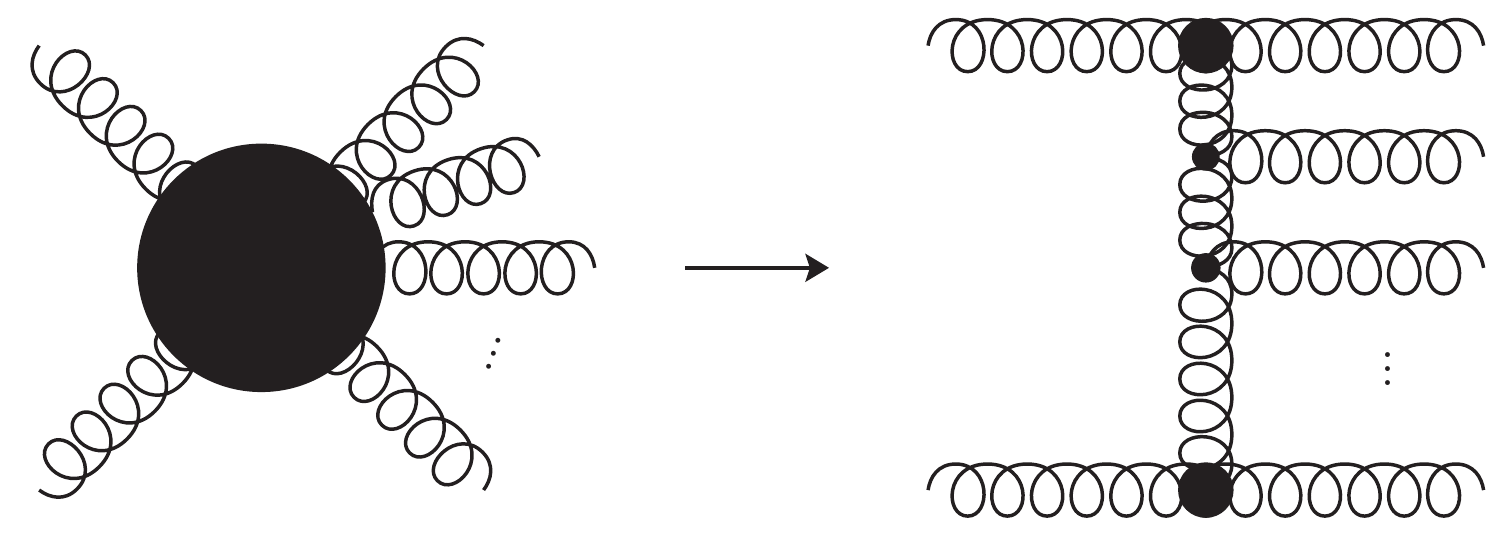}
\caption{Left: Comparison between the full, MRK and HEJ \emph{leading order} matrix elements. Right: Pictorial representation of the HEJ matrix element. }
\label{fig:lo}
\end{figure}

The inclusion of Higgs emission is done by replacing $||S_{XY\to XY}||^2$ with \cite{Andersen:2009nu,hjets}
\begin{equation}
||S^H_{XY\to XY}||^2=\sum_{\rm{helicities}}\left\vert j_\mu(p_1,p_a)\frac{g^{\mu\rho}}{q_{H1}^2}V^H_{\rho\sigma}(q_{H1},q_{H2})\frac{g^{\sigma\nu}}{q_{H2}^2}j_\nu(p_n,p_b) \right\vert^2,
\end{equation}
where
\begin{equation}
V_H^{\mu\nu}(p,q)=\frac{\alpha_s}{6\pi v}(g^{\mu\nu}p\cdot q-p^\nu q^\mu)
\end{equation}
is the effective gluon-gluon-Higgs coupling in the infinite top-mass limit. The $q_{H1}$ and $q_{H2}$ are the t-channel momenta on either side of the Higgs in the rapidity ordered chain. In the left panel of Fig. \ref{fig:lo}, the full $qQ\to qghQ$ LO matrix element is compared to LO MRK and LO HEJ matrix elements, substantiating the effect from the modelling of sub-leading contributions included in HEJ.  

Outside the high energy limit, the final state configurations other than those captured by the Eq. \ref{eq:m} do not vanish anymore. In HEJ these processes are added at tree level accuracy up to four jets. Furthermore the resummed configurations are matched to tree level accuracy up to four jets. 

\begin{figure}[h]
\centering
\includegraphics[width=0.5\textwidth]{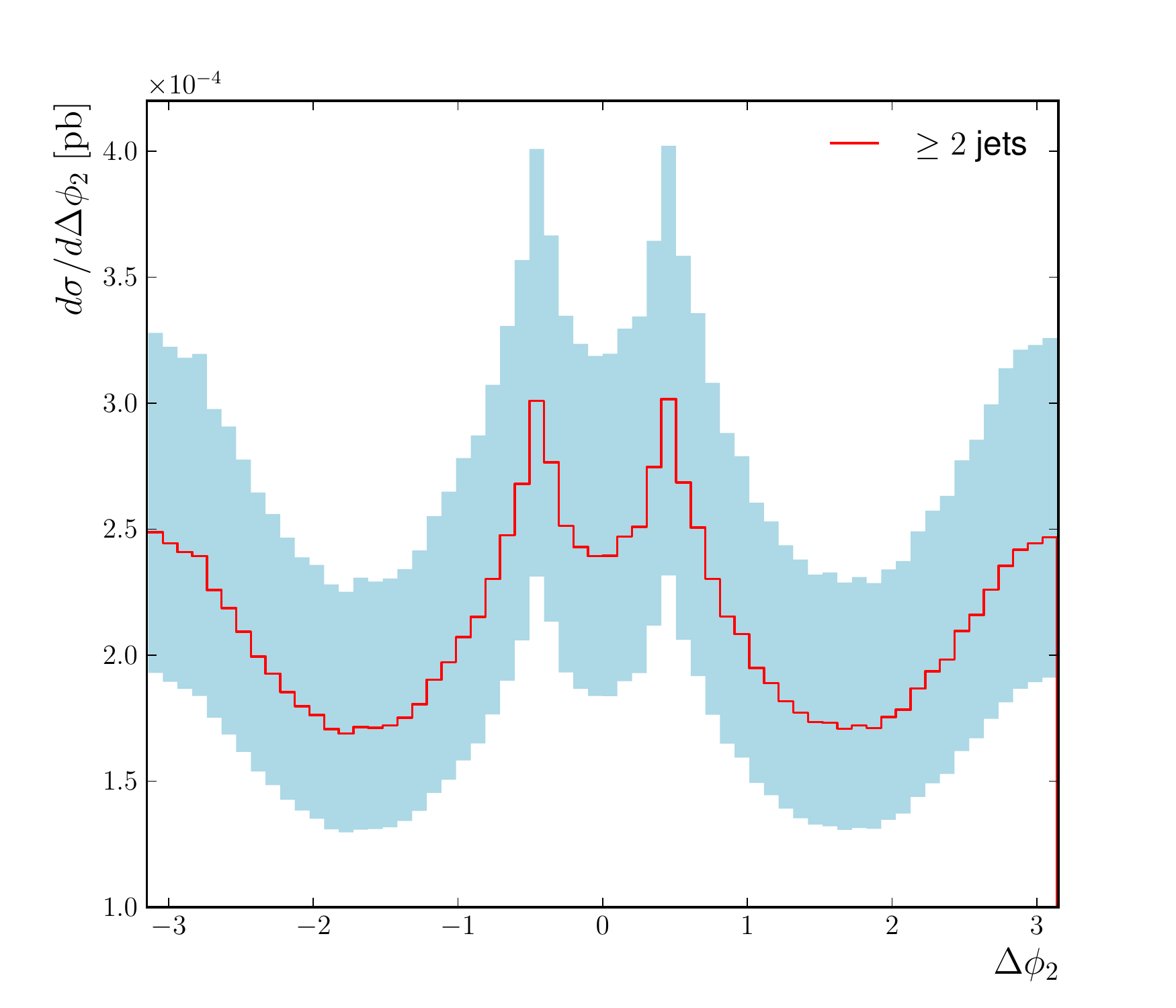}
\caption{Differential cross section as a function of $\Delta \phi_2$. }
\label{fig:phi2}
\end{figure}

Sample plot for the production of a Higgs boson with at least two jets are shown in Fig. \ref{fig:phi2}. The variable $\phi_2$ is the azimuthal angle  between the vector sum of jets forward and jets backward of the Higgs boson. The shape of the distribution is sensitive to the CP-structure of the ggH-coupling \cite{Andersen:2010zx}. The two peaks in the middle are due to would be collinear divergences regulated by the jet definition.

\section{Comparisons}

\begin{figure}[h]
\centering
\includegraphics[width=0.45\textwidth]{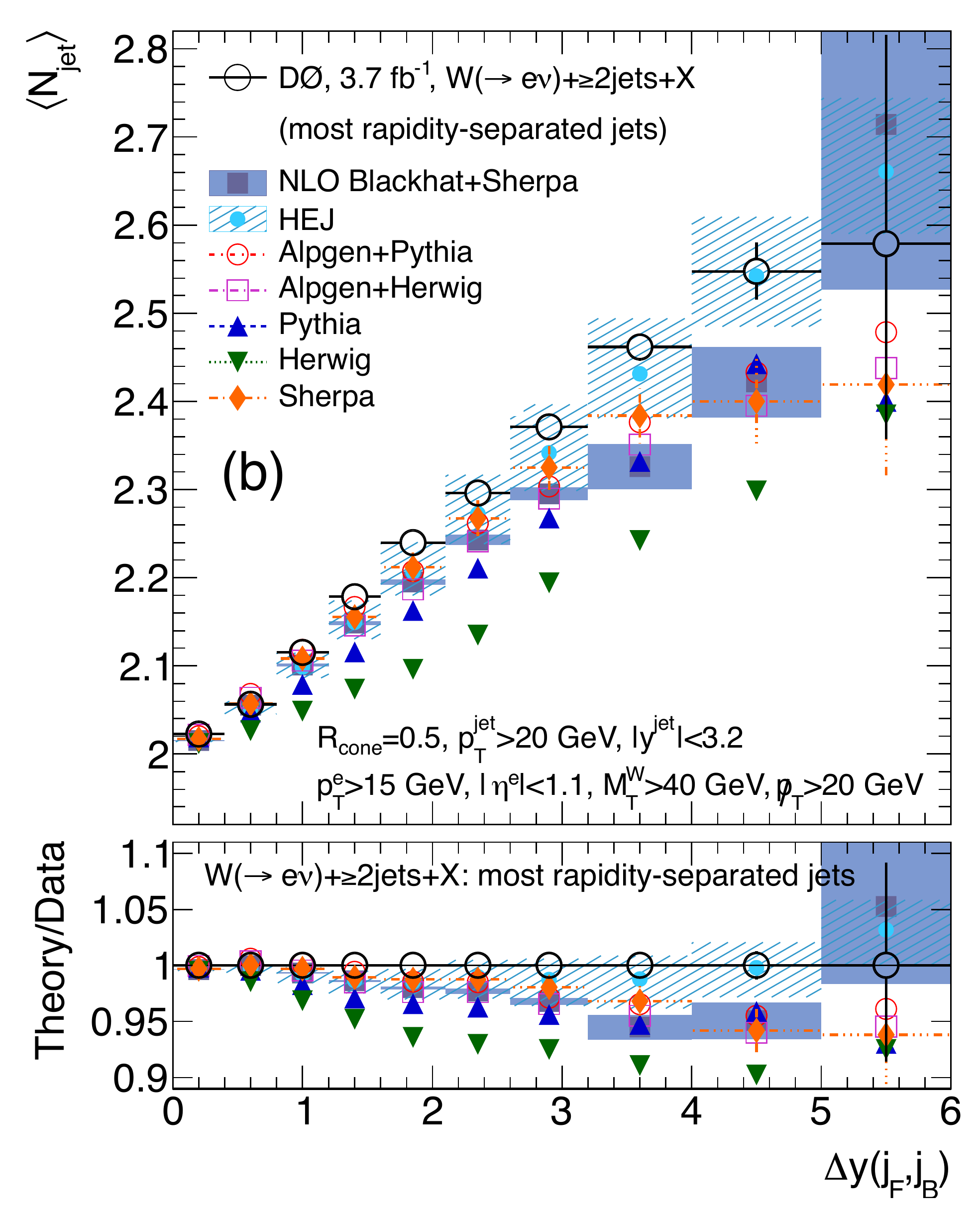}
\includegraphics[width=0.45\textwidth]{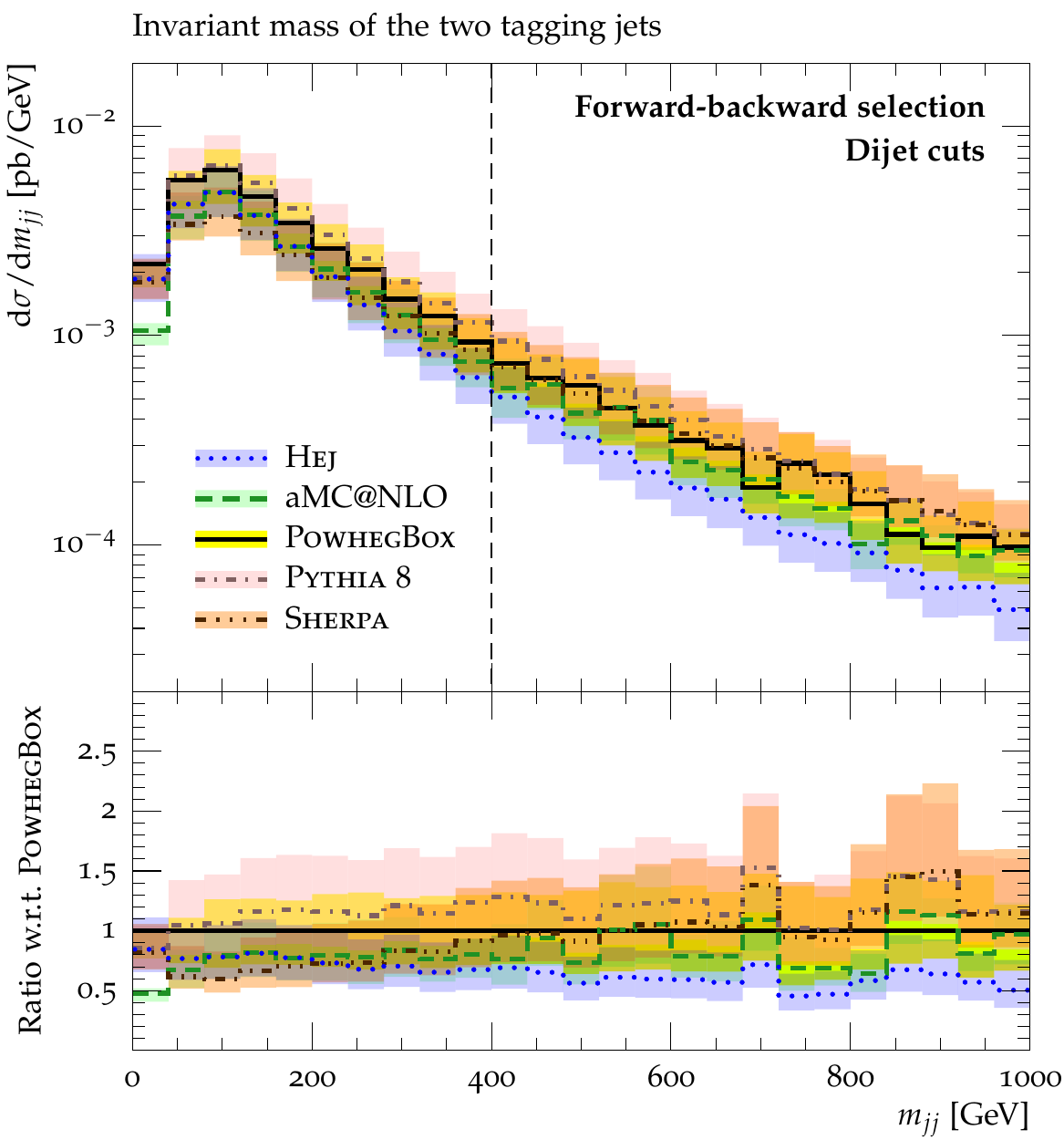}
\caption{Left: The average number of jets as a function of the rapidity difference between the most forward and backward jet. Right: Cross section as a function of the invariant mass of the most forward and backward jets}
\label{fig:d0}
\end{figure}

As was mentioned in the introduction, the Higgs boson production in association with more than two jets is expected to have a similar radiation pattern as the W boson production in association with more than two jets. There is plenty of data for W plus multiple jets which can be used to test the approximation taken in HEJ. The extension to HEJ framework to include the production of W boson is described in \cite{Andersen:2012gk}. In the left panel of Fig. \ref{fig:d0} HEJ is compared against the D0 data \cite{Abazov:2013gpa} with other generators with different approaches. The figure shows the average number of jets as a function of the rapidity separation between the most forward and backward jet. This variable is relevant for jet veto studies. The logarithms included in HEJ becomes more important with increasing rapidity separation. A good agreement between HEJ and data is achieved over the whole range of rapidities. 

 Recently different generators describing the Higgs boson plus multi-jet production were compared in the Les Houches SM working group report\cite{Butterworth:2014efa}. In the right panel of Fig. \ref{fig:d0}  is the differential cross section as a function of the invariant mass of the most forward and backward jets. The description given by HEJ agrees with the other generators at low invariant masses. The difference seen at large invariant masses is expected since the logarithms included in HEJ become important, but are not systematically included in the other frameworks.

The data from LHC will soon allow to test the different descriptions of the Higgs boson plus multi-jet processes. Further work is needed to understand the differences between the descriptions and extract the maximum amount of information from the data.

\section*{Acknowledgments}

This work was supported by the Research Executive Agency (REA) of the European Union under
the Grant Agreement number PITN-GA-2010-264564 (LHCPhenoNet)


\section*{References}

\begin{thebibliography}{99}

\bibitem{Andersen:2007mp} 
  J.~R.~Andersen, T.~Binoth, G.~Heinrich and J.~M.~Smillie,
  JHEP {\bf 0802}, 057 (2008).

\bibitem{Andersen:2009he} 
  J.~R.~Andersen and J.~M.~Smillie,
  Phys.\ Rev.\ D {\bf 81}, 114021 (2010).

\bibitem{Andersen:2009nu} 
  J.~R.~Andersen and J.~M.~Smillie,
  JHEP {\bf 1001}, 039 (2010).
 
  
\bibitem{Andersen:2011hs} 
  J.~R.~Andersen and J.~M.~Smillie,
  JHEP {\bf 1106}, 010 (2011).
  
\bibitem{hjets} 
  J.~R.~Andersen, T.~Hapola and J.~M.~Smillie,
  in preparation.  
  
\bibitem{Andersen:2012gk} 
  J.~R.~Andersen, T.~Hapola and J.~M.~Smillie,
  JHEP {\bf 1209}, 047 (2012).

\bibitem{Butterworth:2014efa} 
  J.~Butterworth {\it et al.},
  arXiv:1405.1067 [hep-ph].
  
\bibitem{Andersen:2010zx} 
  J.~R.~Andersen, K.~Arnold and D.~Zeppenfeld,
  JHEP {\bf 1006}, 091 (2010).
  
\bibitem{Abazov:2013gpa} 
  V.~M.~Abazov {\it et al.}  [D0 Collaboration],
  Phys.\ Rev.\ D {\bf 88}, 092001 (2013).

\end{thebibliography}
\end{document}